# Photovoltaic Self-Powered Gas Sensing: A Review

Xiao-Long Liu, Yang Zhao, Wen-Jing Wang, Sheng-Xiang Ma, Xi-Jing Ning, Li Zhao, and Jun Zhuang

***Abstract*—The self-powered sensing system could harness ambient energy to power the sensor without the need for external electrical energy. Recently, the concept of photovoltaic (PV) self-powered gas sensing has aroused wider attentions due to room-temperature operation, low power consumption, small size and potential applications. The PV self-powered gas sensors integrate the photovoltaic effects and the gas sensing function into a single chip, which could truly achieve the goal of zero power consumption for an independent gas sensing device. As an emerging concept, the PV self-powered gas sensing has been achieved by using different strategies, including integrated gas sensor and solar cell, integrated light filter and solar cell, gas-sensitive heterojunction photovoltaics, and gas-sensitive lateral photovoltaics, respectively. The purpose of this review is to summarize recent advances of PV self-powered gas sensing and also remark on the directions for future research in this topic.**

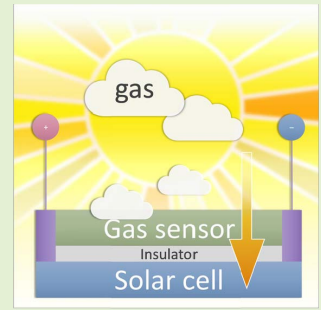

## I. INTRODUCTION

**T**HE self-powered sensing system has become an important research focus due to the growing energy crisis and environmental problems. It could harness ambient energy such as sunlight, lamplight, body motion, and heat, to power the sensor without external electrical energy [1]–[3]. The idea of nanogenerators took a huge step forward towards the integrated and miniaturized self-powered active sensing system combining nanomaterials and microelectronics processing [4]–[10]. The triboelectric nanogenerator (TENG) [5]–[7]

Manuscript received September 16, 2020; accepted November 9, 2020. Date of publication November 11, 2020; date of current version February 5, 2021. This work was supported in part by the National Natural Science Foundation of China under Grant 61675045, in part by the National Basic Research Program of China (973 Program) under Grant 2012CB934200, and in part by the Specialized Research Fund for the Doctoral Program of Higher Education under Grant 20130071110018. The associate editor coordinating the review of this article and approving it for publication was Prof. Minhee Yun. *(Corresponding authors: Jun Zhuang; Li Zhao.)*

Xiao-Long Liu, Yang Zhao, Wen-Jing Wang, Sheng-Xiang Ma, and Jun Zhuang are with the Shanghai Ultra-Precision Optical Manufacturing Engineering Center, Department of Optical Science and Engineering, Fudan University, Shanghai 200433, China (e-mail: junzhuang@fudan.edu.cn).

Xi-Jing Ning is with the Applied Ion Beam Physics Laboratory, Institute of Modern Physics, Department of Nuclear Science and Technology, Fudan University, Shanghai 200433, China.

Li Zhao is with the State Key Laboratory of Surface Physics, Department of Physics, Fudan University, Shanghai 200433, China, and also with the Collaborative Innovation Center of Advanced Microstructures, Fudan University, Shanghai 200433, China (e-mail: lizhao@fudan.edu.cn).

Digital Object Identifier 10.1109/JSEN.2020.3037463

and the piezoelectric nanogenerator (PENG) [8]–[10], for example, could exploit random mechanical energy into the powering source. Recent works have shown the promising concept in the potential applications of wireless or flexible electronic devices [11]–[15], internet of things (IoTs) [16]–[18], and healthcare or medical areas [19]–[22]. Besides nanogenerators, there are other self-powered sensing concepts based on various mechanisms like the photovoltaic (PV) [15], [23]–[25], thermoelectric/pyroelectric [23], [26]–[29], electromagnetic [30], [31], and electrochemical effects [32], [33].

The renewed attention to environmental problems has also driven the need for gas sensing devices, which could identify the type and concentration of gaseous chemicals in the ambient air [34]–[48]. Electrically driven gas sensors based on semiconductors with the most general applicability are widely used in the areas of industry, agriculture, environment, medicine and so forth. Nowadays, introducing micro-electro-mechanical systems (MEMS) into the gas sensing facilitates the development in high-end markets [49], [50]. Although improving the "3S" (sensitivity, selectivity and stability) for gas sensors is a relentless pursuit, there are always other features that determine their areas of application [34], [35]. For example, high working temperature is needed to activate the gas sensing properties for many semiconducting materials especially like metal oxides, whereas it is also accompanied with many disadvantages [36]–[44]. First, grain boundary segregation and interdiffusion effects at high temperatures could shift the long-term performance and cause stability issues [51]. Second, the maintenance of a high temperature leads to increased manufacturing cost and poor reliability [52]. Last but arguably the most importantly, the external heating unit in a gas sensor produces considerable power consumption that limits its further application in new areas [35], [37], [41], [43], [46]–[48]. The addition of noble metals such as Au, Pd, Pt, and Ag as catalysts on the surface may lower the activation energy and also the required temperature [53], but working at ambient temperature should be ideal in most cases. Therefore, there are active demands for novel gas sensing technologies to address these issues.

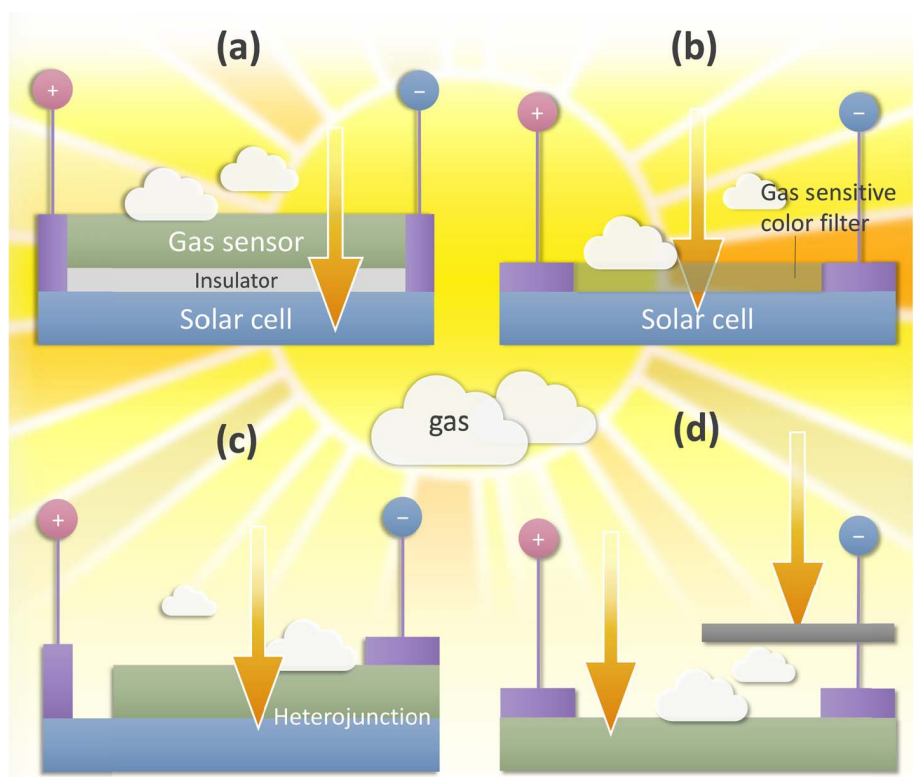

Fig. 1. Schematic diagrams of the different approaches towards photovoltaic self-powered gas sensing. (a) Integrated gas sensor and solar cell. (d) Integrated light filter and solar cell. (c) Gas sensitive heterojunction photovoltaics. (d) Gas-sensitive lateral photovoltaics.

Since the end of last century, it has been established that the gas sensing ability could be activated for many metal oxides under UV light illumination and modulated by light intensity at room temperature instead of thermal activation [54]–[59]. The wavelength of the applied light was gradually replaced by the visible light for some low-dimensional nanoscale semiconductors with narrower bandgaps [60]–[75], to further reduce the power consumption and to avoid the potential harm of the UV light. Later, it was realized that some gas sensors with well-designed structures could be self-sustained under light illumination by using the PV effects even without an external electrical source [76]–[101], which truly achieved the goal of zero power consumption for an independent gas sensing device. Although the PV effects have been long-standing techniques and proven efficient, clean and safe, the concept of the PV self-powered gas sensing has only aroused wider attentions in the last decade and will certainly gain momentum in the future.

The purpose of this review is to summarize recent advances of PV self-powered gas sensing and also remark on the directions for future research in this topic. As an emerging concept, the PV self-powered gas sensing has been achieved by using different strategies. In view of different sensing principles, we mainly identify four types of PV self-powered sensing strategies including integrated gas sensor and solar cell [76]–[78], integrated light filter and solar cell [79]–[83], gas-sensitive heterojunction photovoltaics [84]–[97], and gas-sensitive lateral photovoltaics [98]–[101], as sketched in Fig. 1a–1d, respectively. In the end, basing on the review of current works, we put forward some prospects in this topic, with which we hope to help practitioners of interests better understand and explore in the area.

## II. DIFFERENT APPROACHES TOWARDS PHOTOVOLTAIC SELF-POWERED GAS SENSING

### A. Integrated Gas Sensor and Solar Cell

The solar cell converts the energy of light directly into the electricity that can be used to power a system. This kind of solar-powered system is very mature and widely used in both large-scale systems such as the satellites [102] and small-scale portable devices [1], [103], [104]. The idea conforms to the self-powered system in a broader sense. However, due to the limited efficiency of solar cells at the stage, the size of the solar panels increases with the demand for the electricity in a system. By contrast, micro-sized solar cells are enough for miniaturized sensing devices that consume only little energy. Following the concept, a gas sensor could be integrated with a solar cell in series to form a PV self-powered gas sensor at the nanoscale.

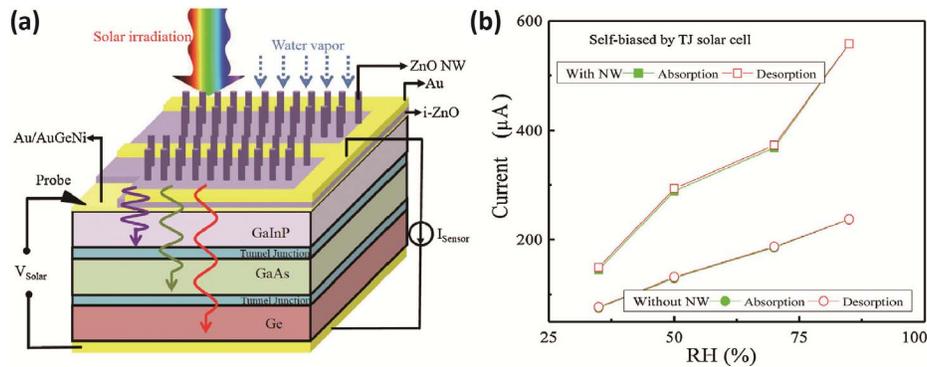

Fig. 2.   ZnO NWs humidity sensor integrated with TJ solar cell [76]. (a) Schematic diagram of the device structure. (b) RH-dependent hysteresis from measured samples showing current responses of the devices with and without ZnO NWs. *Reproduced with permission: 2014, Elsevier.*

As an earlier proof of concept, Hou *et al.* [76] presented an integrated humidity (the content of water vapor in air) sensor and solar cell. For the solar cell, a GaInP/GaAs/Ge triple junction (TJ) was formed and a thin Au/AuGeNi (2 nm/5 nm) layer was deposited as the top electrode, whereas for the humidity sensor, the Au/ZnO Schottky diode was deposited on the Au/AuGeNi layer. To enhance the humidity sensitivity, ZnO nanowires (NWs) was grown atop the structure to increase the surface-to-volume ratio. Fig. 2a shows the schematic diagram of the ZnO NW humidity sensor integrated with TJ solar cell. Under a standard solar irradiation of AM 1.5G (light intensity of 100 mW·cm$^{-2}$), the TJ solar cell produced the open-circuit voltage ($V_{oc}$) of 2.5 V at room temperature, which was used as a self-bias to power the humidity sensor. The humidity-sensing mechanism was proposed to explain the surface conductivity change in the presence of water vapor, indicating an increase in current under a constant bias voltage for the Au/ZnO Schottky structure. The current measurement showed that the response of the self-biased humidity sensor could be enhanced from 75.9, 123.8, 181.2, and 222.6 $\mu$A (without ZnO NWs) to 141.2, 268.1, 365.7, and 572.5 $\mu$A (with ZnO NWs) for 35, 50, 70, and 85% RH, respectively. In the meantime, hysteresis effect showed the current signals could fully recover (Fig. 2b). Transient responses under periodic conditions of 35% and 85% RH showed that the response time was 31 s and 53 s for the sensor with and without ZnO NWs, respectively. This work gives a general idea of the integrated gas sensor and solar cell.

Tanuma and Sugiyama [77] fabricated an integrated CO$_2$ sensor and solar cell based on the similar idea. The solar cell comprised a p-NiO:Li/n-ZnO heterojunction structure, and as for CO$_2$ gas sensing, thin layers of LaOCl/SnO$_2$ were deposited. The sensing layers and the solar cell layers were insulated by SiO$_2$ and connected in series by Au electrodes. Under AM 1.5 irradiation, the integrated devices exhibited the $V_{oc}$ of 0.16 V and short-circuit current density ($J_{sc}$) of 1.04 nA·cm$^{-2}$. Although the resistance response to CO$_2$ gas was observed, the results were rather preliminary and some optimizations were expected such as improving the solar cell performance.

Owing to the development of MEMS technology [49], [50], multiple sensors could be integrated into a single microchip to perform different functions. Juan *et al.* [78] fabricated a self-powered humidity sensor and photodetector (PD) by integrating a crystalline Si interdigitated back-contact (IBC) PV cell with WO$_3$ and Ga$_2$O$_3$ thin films in parallel. The measured $V_{oc}$, $J_{sc}$ and conversion efficiency ($\eta$) of the IBC PV cell under AM 1.5G (100 mW·cm$^{-2}$) illumination were 0.598 V, 35.58 mA·cm$^{-2}$ and 10.29%, respectively. Both photodetection and humidity sensing could work under the constant solar illumination. Regarding photodetection, the cutoff wavelengths of WO$_3$ and Ga$_2$O$_3$ thin films were 370 and 250 nm, respectively, which could be applied to the visible-blind and solar-blind dual band PDs. As for the humidity sensing, the measured current of the WO$_3$ thin film increased monotonically with increased RH.

It can be concluded that this approach of integration towards the PV self-powered gas sensing was mostly based on (top) gas sensor/insulator/(bottom) solar cell tandem structures (expressed by Fig. 1a). As would affect the efficiency of electricity generation, both the gas sensor and insulator layers should be thin or transparent enough, so that the light illumination could penetrate through them into the solar cell. It is also worth mentioning that there are also plenty of well-characterized photovoltaic [105] and gas sensing [34]–[48] materials readily to be integrated using micromachining processes. The groundwork should be the design and optimization of the integrated structures, and on this basis, more functions and better performances could be achieved in the future.

## B. Integrated Light Filter and Solar Cell

For PV devices, the generating capacity directly depends on the amount of the incident light. By using this property, gas adsorption on certain material will change its optical transmittance, thereupon the solar cell beneath the material (*i.e.* a light/color filter) receives different light intensity and changes the electric output. This approach is referred to as the integration of gas sensitive light filter and solar cell (expressed by Fig. 1b).

Kang *et al.* [79], [81] first developed a PV self-powered gas sensor consisting of a colorimetric film and an organic solar cell. The redox reaction of the N,N,N′,N′ -tetramethyl-p-phenylenediamine (TMPD) with NO$_2$ gas could change the color of the TMPD film [106]. To result in more intense color change, a micro-post array pattern of the TMPD was molded using transparent polydimethylsiloxane (PDMS) films. The transmittance of the TMPD coated PDMS film with $NO_2$ exposure decreased in the region of 550 to 660 nm due to color change, and films with larger aspect ratios of micro-post arrays exhibited more decrease in film transmittance. Therefore, the current change of the solar cell below the filter coating reflected the concentration of $NO_2$ gas under the simulated sunlight, although the sensing parameters are not ideal for practical purposes. This means that the sensing mechanism relied mostly on the reaction kinetics of the TMPD film with $NO_2$ [106].

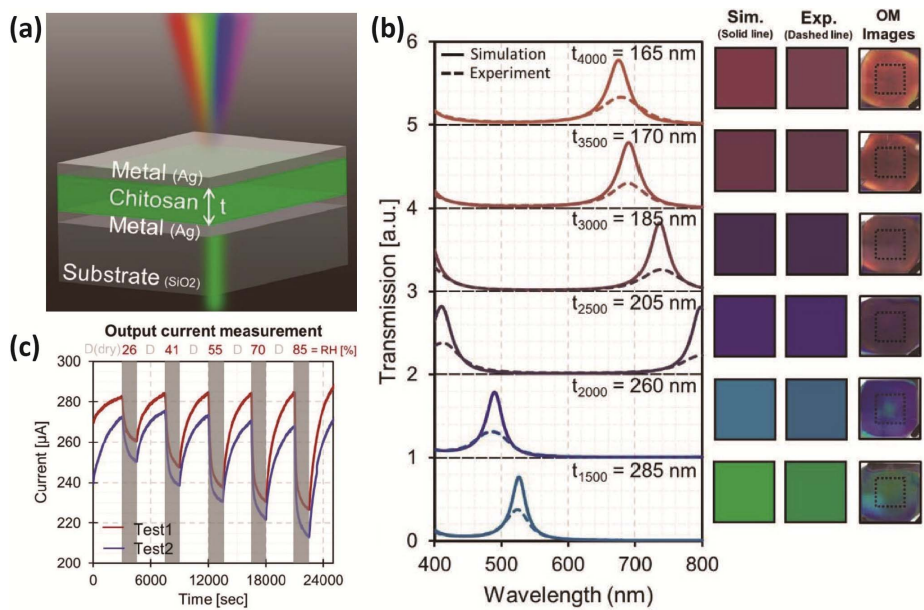

Fig. 3. Self-powered humidity sensor using Chitosan-based plasmonic MIM filters [82]. (a) Schematics of the proposed MIM filter that had thickness *t* of chitosan. (b) Transmission spectra calculated using TMM and obtained using UV–vis/IR spectroscopy measurement of MIM filters having estimated thicknesses of chitosan $t_v$ where *v* represents spin-coating speed. Insets: colors converted from simulation and measured spectra, and optical microscopy images. (c) Output current generated by the proposed humidity sensor under varied RH condition. *Reproduced with permission: 2020, John Wiley and Sons.*

Similarly, Chen *et al.* [80] designed a Pd-$WO_3$/graphene/Si tandem structure with Pd-$WO_3$ as the sensing layer whose transmittance could decrease when being exposed to the hydrogen gas. The transmittance decrement reached the maximum of 72% at the wavelength of ~1000 nm in 4 vol% $H_2$/Ar. The photocurrent decreased by 90% within 13 seconds in 4 vol% $H_2$/Ar under a 980 nm laser irradiation with a power of 10 mW, and exhibited a limit of detection (LOD) of 0.05 vol% with a power of 100 W. The decrease in transmittance of the Pd-$WO_3$ was due to the generation of $H_xWO_3$ ($x = 0$–1) in $H_2$ exposure.

Beyond the chemical method, the transmittance can be modulated by changing the physical structure of the light filter with gas adsorption, based on various filter structures. Jang *et al.* [82] proposed a tunable structural light filter with metal–insulator–metal (MIM) multilayer for humidity detection. The MIM structure worked as a Fabry–Pérot resonator, for which the top and bottom 25 nm Ag layers served as highly reflecting surfaces and chitosan served as the cavity

(Fig. 3a). The key element was the chitosan hydrogel cavity, in which the effective optical thickness $t_{eff}$ and refractive index $n_c$ changed in response to RH. As the thickness of the chitosan layer increased, the transmission peaks red-shifted according to both Transfer-Matrix Method (TMM) simulation and experimental results (Fig. 3b). For self-powered humidity sensing, the MIM filter was combined with a P3HT PV cell. The MIM film used had an initial resonance wavelength $\lambda_{res} = 600$ nm, and the absorption spectrum of the P3HT PV cell drastically decreased at wavelengths $600 \leq \lambda \leq 700$ nm. Therefore, as RH increased, the chitosan layer started to swell, so $\lambda_{res}$ redshift-red and the output current of the P3HT PV cell decreased under light illumination (Fig. 3c). It is worth noting that the current could fully regain its original state (RH = 0%) during the recovery period within 3000 s.

Seo *el al.* [83] also proposed a transmittance changeable filter under the surrounding gas using structural change. The authors fabricated a Pd-polymer nano-transducer with grating structures as the light filter, which was integrated with a solar cell used as the self-powered gas sensor under the visible light. When the Pd was exposed to $H_2$, the generated hydrogenation (Pd$H_x$) with a high mechanical stress caused the deformation of the nano-transducer [107]. Therefore, under the light illumination, the transmittance of the nano-transducer was varied by the $H_2$ gas, and the output electrical signal from the solar cell also changed. The proposed device demonstrated a sensitivity ($\Delta I/I_0$) of 3.1% and response time of 111 s at 2% $H_2$. More importantly, the sensor had satisfactory long-term durability. During >150 cycles of varying $H_2$ concentration exposure from 2 to 0.5%, the device showed reliable and reproducible response curves at all time.

For this approach, the change in optical transmittance induced by gas adsorption could be based on very different mechanisms, depending on diverse types of light filters. Developing light filters with fast response kinetics, greater degree of the transmittance changes and selective response to specific gases are highly desired for the PV self-powered gas sensing. For that matter, the choice of the solar cell should consider matching the wavelengths of the transmittance change for each filter.

### C. Gas-Sensitive Heterojunction Photovoltaics

In most photovoltaic applications, the generation of voltage and current relies on the structure of semiconductor p–n junctions. Basically, the diffusion of charge carriers to the opposite directions creates a space charge region near the p–n junction, where a built-in electric field is established, by which the photogenerated electron-hole pairs are separated. The heterojunction, a junction formed between two dissimilar semiconductors [108], is used not only in solar cells, but also in various device applications especially for gas sensors [24], [34], [38]–[40], [44]. For semiconductor gas sensors, the synthesis of heterojunction nanomaterials (heterostructures) has been extensively studied to exploit the sensing properties arising from the junctions and the effects of hierarchical organizations of dissimilar materials [34], [38]–[40], [44]. Since the heterojunction structure enables both the PV effect and gas sensing, the two functions can be achieved at the same time on a single heterojunction by design, which invokes the approach of the gas-sensitive heterojunction photovoltaics (expressed by Fig. 1c).

It is well known that silicon is the most important material in PV industry, and it has also attracted widespread attention as a gas sensing material, e.g., in the form of porous silicon (PSi) [109], [110]. The PSi has several unique features like large surface area, high chemical activity, visible luminescence and so on. Besides, many physical parameters of the PSi are proven to be sensitive to gas surroundings which are the basis for gas sensor design [109]. Vashpanov et al. [84] first implemented the concept of PV self-powered gas sensing using PSi/Si heterojunction, which was fabricated by forming a thin (12 μm) layer of PSi on the silicon wafer by electrochemical etching in HF-based solutions. The photo-electromotive force (photo-EMF, i.e., photovoltage) of the PSi/Si heterojunction was measured at room temperature under the light illumination as the sensing signal, and it was claimed to be sensitive to ammonia [84], ethanol/mercaptoethanol [88], and acetone [97]. In one of the works, the dependences of the photo-EMF on the light wavelength, the light intensity (characterized by the illumination level) and the NH₃ concentration were preliminarily investigated. Although a simplified model of heterojunction energy diagram was proposed, the effect of the gas adsorption on the change of the photo-EMF was poorly understood.

Hoffmann et al. [85] designed a CdS@n-ZnO/p-Si heterojunction which they referred to as the solar diode sensor (SDS) integrating the functions of both gas sensing and solar energy harvesting (Fig. 4a). The sensor worked by the change of the open-circuit voltage ($\Delta V_{oc}$) under simulated solar (AM1.5) illumination without the external electric power. The $\Delta V_{oc}$ showed opposite directions in responses to oxidizing (oxygen) and reducing gases (methane and ethanol) (Fig. 4b–4c). By comparative experiment, it was confirmed that CdS@n-ZnO (CdS decorated n-ZnO nanorods) acted as gas sensing units and the n-ZnO/p-Si junction formed as the build-in signal generation unit under the light. More importantly, the authors proposed a convincing step-by-step mechanism interpreting the $\Delta V_{oc}$ (Fig. 4d–4i). In combination with corresponding I–V characteristics, the mechanism answered some basic questions: (i) how $V_{oc}$ was generated in the structure under the solar illumination, and how (ii) oxidizing and (iii) reducing gases changed the $V_{oc}$ with opposite directions. Theoretically, the semiconductor quantum theory gives a qualitative description of the mechanism [108], with the build-in potential ($V_{bi}$) in the n-ZnO/p-Si junction expressed as

$$V_{bi} = \frac{kT}{q} \ln\left(\frac{N_D^{ZnO} N_A^{Si}}{N_i^{ZnO} N_i^{Si}}\right) \tag{1}$$

in which $N_D$, $N_A$ and $N_i$ are donor, acceptor and intrinsic doping level; $k$, $T$ and $q$ are Boltzmann constant, absolute temperature and electron charge, respectively. The interaction of oxidizing (reducing) gases with CdS@n-ZnO caused a decrease (increase) in the donor carriers $N_D$ in the ZnO and thus a decrease (increase) in the build-in potential according to (1).

Based on the above work, Hoffmann et al. [86] and Gad et al. [87] fabricated a highly selective and sensitive self-powered NO₂ gas sensor using the nanostructured n-ZnO/p-Si heterojunctions. To improve the sensitivity, 16 nanodiodes were connected in series by design which led to a high $V_{oc}$ output of 1.35 V under simulated solar (AM1.5). As for selective gas sensing, the surface of the n-ZnO NWs was modified by a self-assembled monolayer (SAM) of organic functionality. Two different methoxysilanes were used to form SAMs on the surface of the n-ZnO NWs, abbr. as amine- and thiol-functionalities respectively. For the amine-functionalized sensor, the detection limit for NO₂ was approximated 170 ppb. More importantly, the response to 0.75 ppm NO₂ was significantly higher than those to 2.5 ppm SO₂, 25 ppm NH₃ and 25 ppm CO. The density functional theory (DFT) calculations indicated the electronic structure of the SAM–NO₂ system depended on the choice of SAM and the molecular orbital energies were altered by NO₂ adsorption, which might determine the selective performance.

Since then, similar structures were developed based on different sensing materials. Jia et al. [89] proposed a PV self-powered NO₂ gas sensor using carbon nanotube membrane/silicon nanowires (CNT/SiNW) heterojunction (Fig. 5). Under cold LED light, the device showed a fast response of $V_{oc}$ of 4–6 s to 10 ppm NO₂ and a significant increase in conversion efficiency ($\eta$) of nearly 200 times to 1000 ppm NO₂. Different from the semiconductor p–n junction, the CNT/SiNW structure was considered as a Schottky junction with the CNT membrane as a metallic material. According to the proposed energy band diagram for

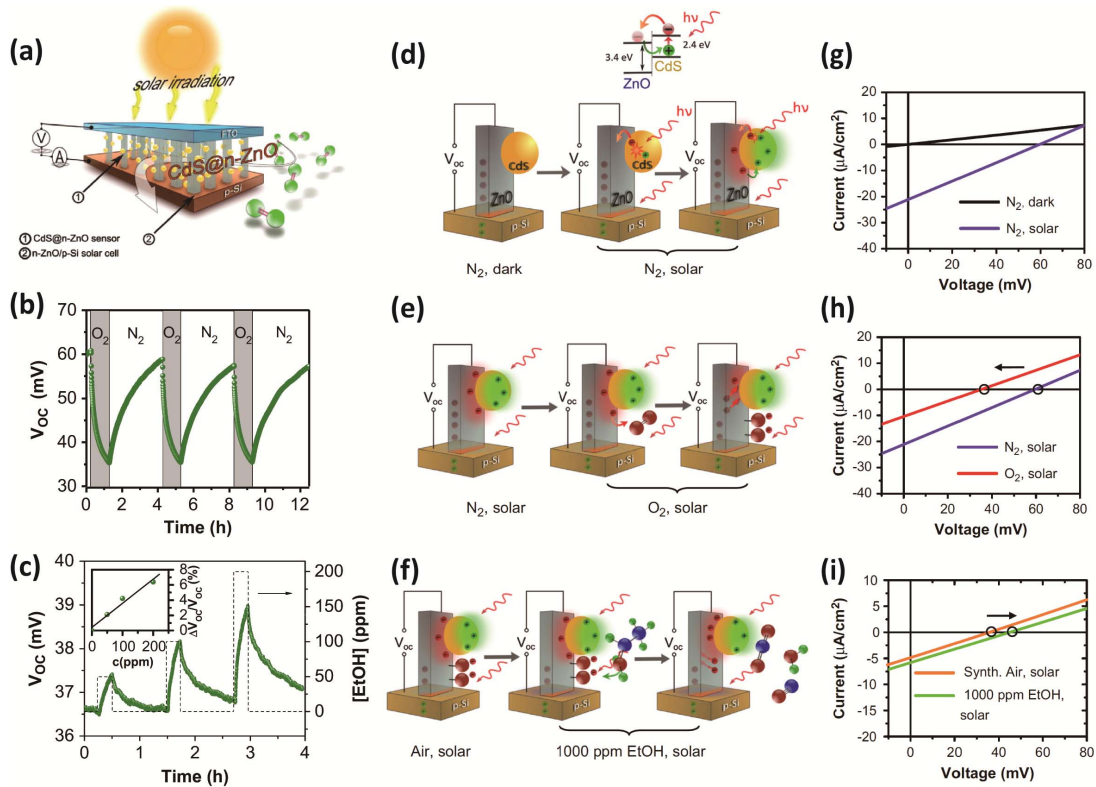

Fig. 4. Solar diode sensor (SDS) based on CdS@n-ZnO/p-Si heterojunction [85]. (a) Schematic diagram of the sensor structure. (b–c) Gas sensing ON/OFF curves in responses to (b) oxidizing (O$_2$/N$_2$) and (c) reducing (EtOH/air) gases recorded with open-circuit voltage ($\Delta V_{OC}$) under simulated solar illumination. (d–i) Proposed step-by-step SDS sensing mechanism. (d) High energy electron patterning of CdS@n-ZnO/p-Si under solar illumination. (e) Charge carrier distribution of CdS@n-ZnO/p-Si in N$_2$ and O$_2$ atomspheres. (f) Charge carrier distribution of CdS@n-ZnO/p-Si in air and EtOH atmospheres. (g–i) Corresponsding I-V curves to (d–f), respectively. *Reproduced with permission: 2013, Elsevier.*

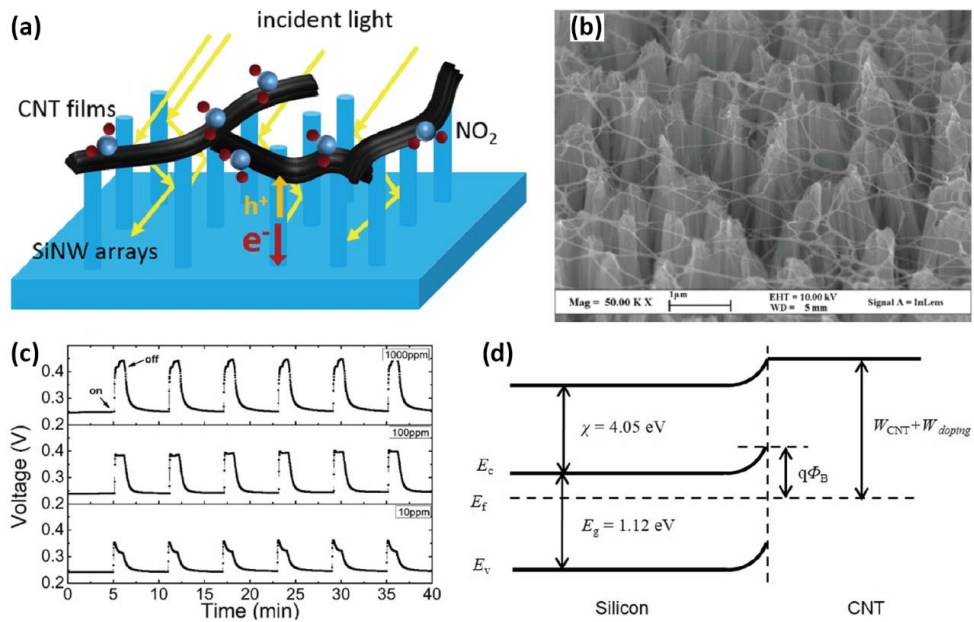

Fig. 5. A photovoltaic self-powered NO$_2$ gas sensor based on CNT/SiNW heterojunction [89]. (a) Schematic diagram of the sensor structure. (b) SEM images of CNT/SiNW heterostructures. (c) Real-time NO$_2$ detection with different concentrations under light illumination. (d) Proposed energy band diagram of CNT/SiNW heterojunction (without the consideration of the band gap broadening of SiNW). *Reproduced with permission: 2016, Springer Nature.*

CNT/SiNW junction (Fig. 5d), the barrier height ($\Phi_B$) on the CNT side is expressed as [89]

$$\Phi_B = (W_{CNT} + W_{doping})/q - \chi \qquad (2)$$

where $q$ is the electron charge, $\chi$ is electron affinity of silicon, $W_{CNT}$ is the work function of CNT membrane and $W_{doping}$ is the additional work function considering gas adsorption as surface doping. Furthermore, the $V_{oc}$ for Schottky junction solar cells can be expressed as

$$V_{oc} = n \left[ q\Phi_B + (kT/q) ln \left( I_s/A_e A^* T^2 \right) \right] \quad (3)$$

where $n$ is the diode ideality factor, $k$ is the Boltzmann constant, $T$ is the working temperature, $I_s$ is the diode saturation current, and $A_e$ and $A^*$ are the contact area of the diode and the Richardson constant, respectively. The adsorption of $NO_2$ caused an increase in the work-function of CNT ($W_{doping}$), which led to an increase in $\Phi_B$ and $V_{oc}$ according to (2) and (3), respectively.

From the above cases, it can be inferred that the change of $V_{oc}$ is positively correlated to the change of barrier height in the heterojunction by gas adsorption. Indeed, there are further examples that could prove the point. Liu *et al.* [90] proposed a PV self-powered gas sensor based on single-walled carbon nanotubes (SWCNT)/Si heterojunction. Under monochromatic light illumination of 600 nm, the $V_{oc}$ decreased in response to $H_2S$, which originated from the decreased holes concentration in SWCNT and thus the decrease in the barrier height. Lee *et al.* [93] proposed a PV self-powered gas sensor based on graphene (Gr)/bulk Si or Gr/$WS_2$ heterojunction. Upon visible light illumination, the short-circuit current ($I_{sc}$) increased in oxidizing gas ($NO_2$) and decreased in reducing gases ($NH_3$ and $H_2$). The Gr was decorated by Pd nanoparticles with catalysis effect so that it had higher (lower) work function in oxidizing (reducing) gases. Liu *et al.* [94] proposed a PV self-powered gas sensor based on p-SiNW/ITO heterojunction. Under visible light illumination, the sensor showed a decrease in the $I_{sc}$ after $NO_2$ adsorption, due to a decrease in electron concentration in the SiNW and thus a decrease in the barrier height.

Apart from silicon, the perovskites family has become one of the most promising materials for the next-generation optoelectronics [111]. Meanwhile, the perovskite semiconductors have also exhibited sensing abilities for chemical species at room temperature [112]. H. Chen *et al.* has developed several PV self-powered gas sensors using different perovskites materials including $CsPbBr_3$ (CPB) [91], $CsPbBr_2I$ (CPBI) [92] and $FA_{0.80}MA_{0.15}Cs_{0.05}PbI_{2.55}Br_{0.45}$ (FMCPIB) [96] perovskites. The authors first used inorganic halide perovskites for self-powered chemical sensing, and CPB was chosen for its stability at ambient condition [91]. The prototype device was based on Au/CPB/FTO structure with a porous network of CPB, and it was designed for detecting medical-related chemicals like $O_2$, acetone and ethanol. The device generated the open-circuit voltage ($V_{oc}$) of 0.87 V and the short-circuit current density ($J_{sc}$) of 1.94 nA·cm$^{-2}$ at room temperature under visible light illumination (AM 1.5, 37.8 mW·cm$^{-2}$). The photocurrent was measured as the sensing signal and the responsivity was defined by the ratio of the change of photocurrent to the basecurrent (*e.g.*, ($I_{O2}$ − $I_{N2}$)/$I_{N2}$). In exposure to pure $O_2$, the responsivity was around 0.93 with a swift response/recovery time of 17/128 s, respectively; the device could easily detect down to 1% $O_2$ in $N_2$. For acetone and ethanol, the device could quickly detect a low concentration of 1 ppm with responsivity of ≈0.03 and 0.025, respectively. Different from the results obtained from other heterojunctions, here, the photocurrent increased in response to both oxidizing and reducing gases. The authors proposed that $O_2$, ethanol or acetone acted as vacancy filler and reversibly fill in intrinsic Br vacancies of the CPB. With the decrease in Br vacancies traps for photoexcited charges, the photocurrent increased with increasing concentration of the analytes.

Then, Chen *et al.* [92] improved the structure by replacing the CPB with $CsPbBr_2I$ (CPBI) as the CPBI has a broader light absorption range than the CPB. They also deposited a hole-blocking layer of $TiO_2$ above the FTO layer, forming Au/CPBI/$TiO_2$/FTO structure. With the improvements, the device generated the $V_{oc}$ of 0.9 V and the $I_{sc}$ of 37 nA under simulated sunlight illumination (AM 1.5, 42.3 mW·cm$^{-2}$). Upon exposure to 4 and 8 ppm of acetone in an inert $N_2$ atmosphere, the photocurrent increased from 8.8 nA to 9.9 nA and 10.2 nA, respectively; a LOD of 1 ppm was determined for both acetone and methanol. However, the responsivity was much lower in the background gas of $O_2$ than in $N_2$. Besides, the sensor also responded to $NO_2$ and other VOCs like propane, ethanol and ethyl-benzene, which indicated that the selectivity required further improvements.

Very recently, Chen *et al.* [96] turned to the triple cation FMCPIB perovskites with the unique features of self-charging and energy-storage. A sensor structure of Carbon/FMCPIB/$TiO_2$/FTO (Fig. 6a and 6b) was fabricated, which generated the $V_{oc}$ of 0.63 V and $I_{sc}$ of 18.6 nA under the fluorescent lamp light irradiation (≈35 $\mu$W·cm$^{-2}$). Interestingly, photoluminescence (PL) characterization of the FMCPIB perovskite films revealed a negligible response to the presence of $O_2$, making them attractive for chemical sensing in atmospheric conditions. Under a weak fluorescent lamp irradiation of ≈1.3 $\mu$W·cm$^{-2}$, the photocurrent increased from 2.95 to 3.35 nA with $NO_2$ concentration increasing from 0 to 8 ppm in air, and the measured response/recovery time was only 16.7/126.1 s, respectively. Furthermore, the selectivity was much improved (Fig. 6c). The sensor demonstrated excellent performance for $NO_2$ detection with an average response of ≈0.08 for 8 ppm of $NO_2$ and negligible responses (<0.014) to exposure of up to 20% $O_2$ + 8% $CO_2$, 8 ppm methanol (MtOH), ethanol (EtOH), and acetone gases. The selective response to $NO_2$ was attributed to the surface amine groups on the FMCPIB. More notably, the light irradiation being switched off, the sensor still exhibited repetitive response to $NO_2$ gas for 1.7 h without any external power (Fig. 6d). A total of ≈0.556 $\mu$C charges were transferred across the electrodes during the period, which revealed a large energy storage capacitance at the $TiO_2$/FMCPIB/carbon interfaces. The finding may be utilized for self-powered sensing in the dark.

In summary, Table I intuitively compares the sensing performance of the PV self-powered gas sensors based on the heterojunction structures discussed in this section mostly. When developing a new gas sensor, one should consider the following steps to fully demonstrate its sensing ability:

1) Identifying the materials or parts corresponding to the sensing function and the light-harness function. In the case of CdS@n-ZnO/p-Si structure [85], the CdS@n-ZnO interacted with the gas molecules and the n-ZnO/p-Si heterojunction generated a build-in potential under light illumination.

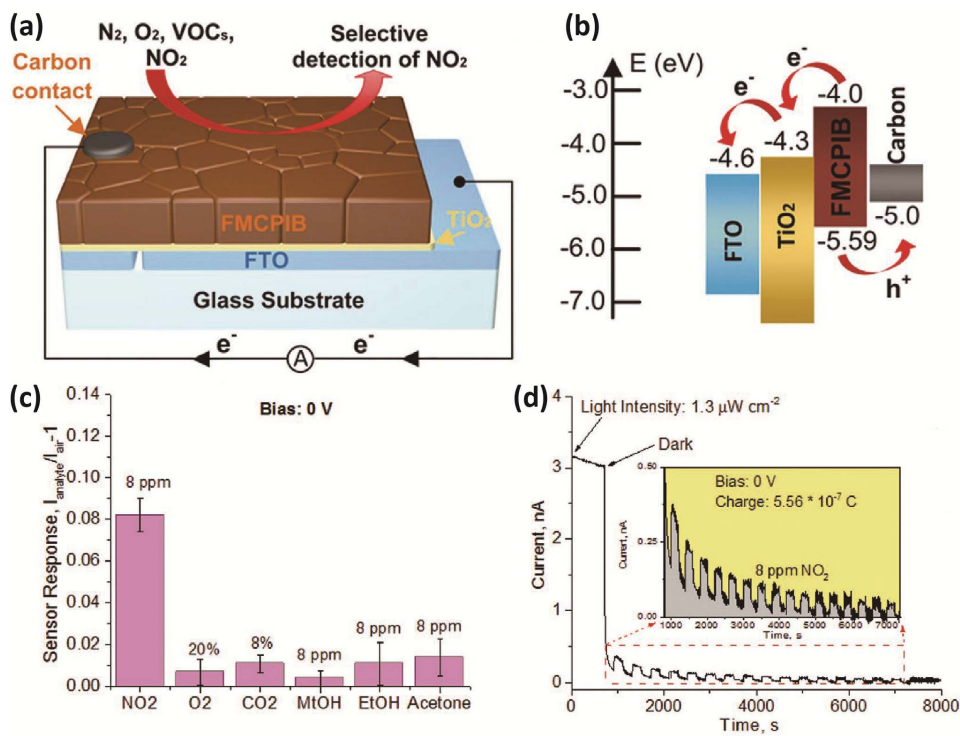

Fig. 6. Photovoltaic self-powered gas sensor based on perovskites semiconductors. Scheme representation of the **(a)** The Carbon/ FMCPIB/TiO₂/FTO structure with **(b)** the corresponding band diagram. **(c)** The sensor response of FMCPIB sensor for six different gases under weak fluorescent lamp light irradiation (1.3 $\mu W \cdot cm^{-2}$). **(d)** Dynamic current response of self-powered FMCPIB devices upon switch-off of fluorescent light irradiation, followed by 17 consecutive injections of 8 ppm of NO₂ in the dark; the inset in **(d)** is the enlarged part of the red dashed region with the integration of the charges. *Reproduced with permission: 2020, John Wiley and Sons [96].*

2) Selecting the appropriate light source. For silicon or perovskites, solar spectrum or visible light LED is ideal for their matched band gaps and the accessibility of the light source. However, a light source with shorter wavelengths may be chosen for wide band gap materials like most of the metal oxides (when considered as light absorber) [95].

3) Characterizing the typical sensing parameters for the target gas, such as sensitivity or gas response, response and recovery time, detection range, selectivity, stability, etc. Ideally, the variation of sensing parameters in different illumination conditions (*e.g.*, light wavelength and intensity) should be studied.

4) Understanding the mechanism of the light to electricity conversion. This always requires the development of the band diagram of the heterojunction structure.

5) Understanding the mechanism of gas-semiconductor interaction. The gas sensing material should be characterized in advance for accurate modelling. The theory of electronics may provide some perspectives on how gas adsorption affects the electronic structure of the material [113], thereby changing the sensing signal.

In some cases of the gas-sensitive heterojunctions, the photovoltaic and gas sensing mechanism can be well-established. Therefore, a further quantitative analysis of the sensing dynamics is highly anticipated, which may provide a validation for the sensing mechanism and a better guidance for the precise and predictable design towards a superior PV self-powered gas sensor.

### D. Gas-Sensitive Lateral Photovoltaics

Different from the p–n/Schottky junction solar cell mechanism, anomalous photovoltaic effects (APE) can be observed in certain semiconductor [114]. Although anomalously large photovoltages up to several thousand volts have been reported [115], materials that exhibit the APE generally have very low power generation efficiencies and are not particularly suitable for power generation. Nevertheless, some materials with the APE are already commercialized in the fields of sensors. For example, the position sensitive device (PSD) can measure a position of a light spot on a sensor surface owing to the lateral photovoltaic effect (LPE). For junction photovoltaics, the photovoltage is generated transversely to the junction boundary; by contrast for the LPE, the photovoltage is parallel to the junction [114], [116]. Peculiarly, the LPE could be generated in a semiconductor film even without a junction structure [117]. This is often ascribed to the photoelectric Dember effect which originally describes the Photo-EMF in a homogeneous limited semiconductor under non-uniform illumination [118]. Dember voltage is generated when carriers diffuse away from the illuminated region while electrons and holes have different mobilities [114].

Recently, the lateral photovoltaic effect found application also in the self-powered gas sensing [98]–[101].

TABLE I
Sensing Performance of the Photovoltaic Self-Powered Gas Sensors Based on
Heterojunction Structures in Recent Publications

| Device structure [a] | Light source | Measured signal (scale) | Target gas (LOD) [b] | Typical response [c] (gas concentration) | Typical $t_{res}/t_{rec}$ ([c]) [d] |
|---|---|---|---|---|---|
| CdS@n-ZnO/p-Si [85] | Simulated sunlight (AM 1.5) | $V_{oc}$ (mV) | $O_2$<br>EtOH (~50 ppm)<br>$CH_4$ (~50 ppm) | ~−0.41 (pure $O_2$/$N_2$)<br>~0.065 (200 ppm EtOH)<br>~0.055 (400 ppm $CH_4$) | ~15/60 min |
| p-Si/n-ZnO series [86] (amine- and thiol-functionalities) | Simulated sunlight (AM 1.5) | $V_{oc}$ (V) | $NO_2$ (170 ppb) | Amine: 0.075(250 ppm)<br>Thiol: −0.098(250 ppm) | >1000 s |
| CNT/SiNW [89] | Cold LED | $V_{oc}$ (mV) | $NO_2$ (10 ppm) | 0.46, 0.63, 0.83<br>(10, 100, 1000 ppm) | 4–6 s (10 ppm) |
| SWCNTs/Si [90] | Visible light (600 nm, 1.8 mW·cm$^{-2}$) | $V_{oc}$ (mV) | $H_2S$ (100 ppb) | 0.0223 (400 ppb) | <54 s |
| Au/CPB/FTO [91] | Simulated sunlight (AM 1.5, 37.8 mW·cm$^{-2}$) | $I_{sc}$ (nA) | $O_2$ (1%)<br>Ac (1 ppm)<br>EtOH (1 ppm) | 0.93 (pure $O_2$)<br>0.03 (1 ppm Ac)<br>0.025 (1 ppm EtOH) | 17/128 s (pure $O_2$);<br>9.8/5.8 s (1 ppm Ac) |
| Au/CPBI/TiO$_2$/FTO [92] | Simulated sunlight (AM 1.5, 42.3 mW·cm$^{-2}$) | $I_{sc}$ (nA) | MtOH (1 ppm)<br>Ac (1 ppm) | ~0.16 (8 ppm Ac) | 100–150 s |
| Gr/Si/metal or Gr/WS$_2$/Gr [93] | While light | $I_{sc}$ (μA) or $V_{oc}$ (mV) | $NO_2$, $NH_3$<br>$H_2$ (1 ppm) | ~0.7 (5 ppm $NO_2$)<br>~0.17 (50 ppm $H_2$) | ~500 s (50 ppm $H_2$) |
| p-SiNW/ITO [94] | White LED (576 nm, 20 mW·cm$^{-2}$) | $I_{sc}$ (μA) | $NO_2$ (5 ppb) | ~0.9 (1 ppm) | 80/850 s (1 ppm) |
| Au@rGO/GaN(NRs)/Si [95] | Monochromatic light (382/516 nm, ~1 mW·cm$^{-2}$) | $I_{sc}$ (μA) | CO (<5 ppm) | ~38% (20 ppm) | / |
| Carbon/FMCPIB/TiO$_2$/FTO [96] | Fluorescent lamp (1.3 μW·cm$^{-2}$) | $I_{sc}$ (nA) | $NO_2$ (1 ppm) | 0.08 (8 ppm) | 17/126 s (8 ppm) |

[a] CNT = Carbon nanotube, SiNW = Silicon nanowire, SWCNTs = Single-walled carbon nanotubes, CPB = CsPbBr$_3$, FTO = Fluorine-doped tin oxide, CPBI = CsPbBr$_2$I, Gr = Graphene, rGO = reduced graphene oxide, NR = Nanorod, FMCPIB = FA$_{0.80}$MA$_{0.15}$Cs$_{0.05}$PbI$_{2.55}$Br$_{0.45}$; [b] LOD = limit of detection, Ac = acetone, EtOH = ethanol, MtOH = methanol; [c] Gas response is defined as $\Delta I/I_0$ or $\Delta V/V_0$ in which $I_0$ and $V_0$ are the short-circuit current and open-circuit voltage in the background gas (air or $N_2$, not differentiated here), respectively; [d] $t_{res}/t_{rec}$ = response/recovery time (The definition slightly differs in works), [c] = gas concentration.

X.-L. Liu *et al.* proposed a self-powered ammonia gas sensor using the sulfur-hyperdoped silicon [99]. Silicon after hyper-doping of impurity atoms (sometimes referred to as the black silicon) was known for its superior optoelectronic properties especially for the significant photoresponsivity in visible-NIR wavelengths [119]. The sulfur-hyperdoped silicon used for the gas sensing was prepared by intense femtosecond-laser irradiation on near-intrinsic silicon wafer in a $SF_6$ atmosphere. The material had nanostructured surface morphology and was rich in sulfur-related defects. It absorbed nearly 100% of the incident light due to the unique material characteristics. The sensor was fabricated by evaporating two coplanar metal electrodes on the laser-structured area as ohmic contacts (Fig. 7a). The change of resistance between the two contacts could be used as the sensing signal (in conductometric sensing mode). However, for the self-powered gas sensing, the authors created an asymmetric light illumination (ALI) method to generate a lateral photovoltage. That is, only half side of the surface was illuminated by the light and the other was kept dark to produce an asymmetric light distribution on the surface purposely (Fig. 1d and 7b). Therefore, the photocurrent was generated under the ALI which could be used as the self-powered sensing signal. The sensor showed linear *I-V* characteristics with or without the light or the gas (Fig. 7c), which was different from the heterojunctions with rectifying effects. Under the self-powered working mode, the sensor exhibited a high response of 77% with exposure to 50 ppm $NH_3$ in air and fast response/recovery time of 55/125 s. In contrast, the sensor's response in the dark was nearly neglectable (Fig. 7d). For the sensing mechanism, Dember effect was used to explain the generation of photocurrent under the ALI. In the longitudinal direction, there was an n$^+$/n junction between the S-hyperdoped silicon and the silicon substrate. The build-in electric field drove the photogenerated holes back to the substrate whereas the photoelectrons remained in the upper layer. The photoelectrons diffused away from the illuminated region in the lateral direction, thereby generating the photocurrent. After $NH_3$ adsorption, the photocurrent increased probably due to the reduced recombination rate of photoelectrons on silicon surface. Therefore, the sensing ability of this self-powered sensor was stimulated by light and directly based on the photoelectrons.

The sensing performance of the sensor based on the hyper-doped silicon can be further improved by flexibly adjusting the operating conditions. For example, Liu *et al.* [98], [99] proposed that the gas response of the sensor based on the hyperdoped silicon could be enhanced by at least two orders of magnitude under optical and electrical dual drives, *i.e.* under the ALI and a small but precise negative bias voltage. Both the gas response and response/recovery time could be enhanced by controlling the light intensity or the bias voltage [99]. The optimization method is especially useful for improving the gas sensing performance of the sensor after substantial aging with deteriorative sensing parameters [100]. Furthermore, the authors discovered that the nitrogen-hyperdoped silicon was ultrasensitive to $NO_2$ gas with a very low LOD of 11 ppb [101]. Combing the conductometric and the self-powered sensing modes, the sensor exhibited distinct $NO_2$ gas sensing performance with an ultrawide dynamic range of nearly 6 decades concentration [101].

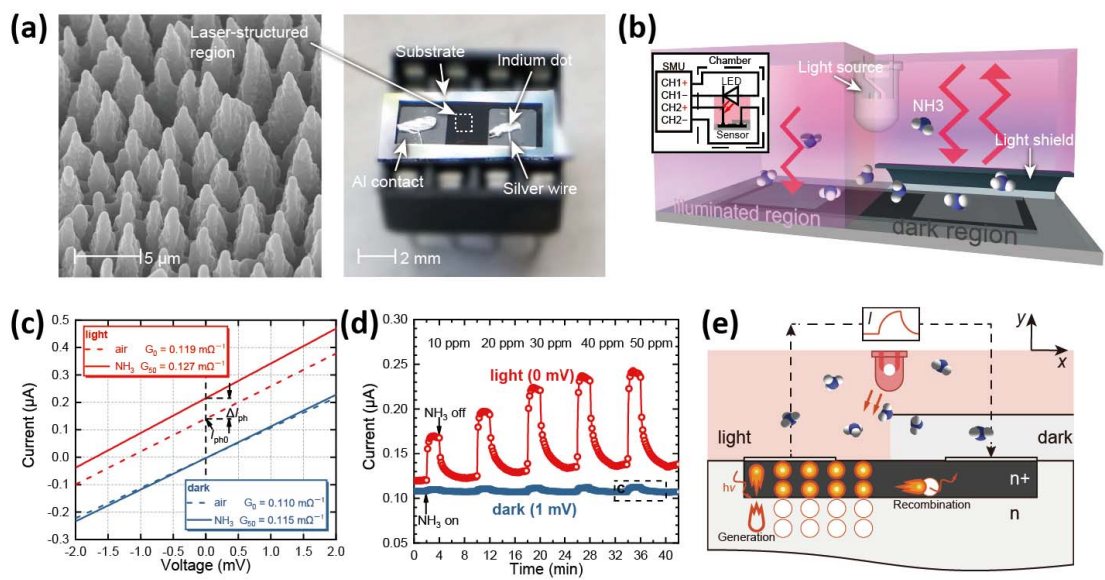

Fig. 7. Self-powered gas sensor based on the sulfur-hyperdoped silicon [99]. (a) The scanning electron micrograph of the laser-structured region and the optical image of the gas sensor with indications of device components. (b) Schematic illustration of the sensor (not to scale) under the asymmetric light illumination (ALI) with inset showing the measurement circuit. (c) Comparison of the responses of I-V characteristics of the sensor in the dark and in the light. (d) Continuous photocurrent responses to 10–50 ppm incremental gas concentrations. (e) Proposed model demonstrating the movement of excess carriers in the material and the origin of the measured photocurrent. *Reproduced with permission: 2019, Elsevier.*

Certainly, the anomalous photovoltaics were also observed in numerous other systems, with different mechanisms arising from varied material and structural characteristics. A combination of the anomalous photovoltaic effects with the gas sensing properties of the materials may result in more unfamiliar physical phenomena. As already indicated, very encouraging gas sensing results have been achieved using simple methods and structures, so it can be worthy of further extensive research in this aspect.

In addition, Table II compares some of the key parameters of PV self-powered gas sensor between different approaches. It should be noted that the high efficiency or high electric output of the solar cell alone does not necessarily guarantee satisfying gas sensing performance. It is more important to have a greater contrast of the optoelectronic signals before and after gas adsorption, which is normally the definition of gas response.

## III. FUTURE DIRECTIONS AND POTENTIAL APPLICATIONS

The concept of the PV self-powered gas sensing is heading towards different directions, for each approach has its own characteristics. This trend will continue, as the development of different gas sensors with various materials, types and principles will be always needed. Thus, there is no single answer to the concept of PV self-powered gas sensing. In this section, we are merely providing some prospects based on the insights from relevant literatures. Although we hope interested readers may find them inspiring, it is highly encouraged that they combine their individual research areas, as multidisciplinary studies are essential from concepts to real applications. The prospects and challenges are summarized from four aspects as follow.

### A. Optimization of Existing Technology

In this review, different approaches towards PV self-powered gas sensors commonly use photovoltaic effects to harvest light energy to power the gas sensor. Therefore, the basic functions of a PV self-powered gas sensor are the powering function and the sensing function. However, they are implemented very differently. According to different sensor concepts, the above approaches can be classified into two major categories. In category I, different material or structural components with only single function are integrated, whereas, in category II, materials or structures with multiple functions are directly used. More specifically, the schematic diagrams of the different approaches, towards PV self-powered gas sensing are shown in Fig. 1a–1b, respectively (viz. A–D in Table II). In the following, the basic sensor concepts of different approaches are summarized and the challenges towards the optimization of existing technology are addressed.

TABLE II
Comparison of Typical Sensing Parameters Between Different PV Self-Powered Gas Sensing Concepts

| Sensor concept | Structure | Light condition | Key PV performance [a] | Key gas sensing performance |
|---|---|---|---|---|
| A. Integrated gas sensor and solar cell | ZnO NW gas sensor with GaInP/GaAs/Ge TJ solar cell [76] | AM 1.5G, 100 mW·cm$^{-2}$ | $V_{OC}$ = 2.5 V, $J_{SC}$ =12.14 mA·cm$^{-2}$ | $I$ from 141 μA@35%RH, to 573 μA@85%RH, $t_{res}$ = 53 s |
| | SnO$_2$ gas sensor with NiO/ZnO solar cell [77] | AM 1.5, 100 mW·cm$^{-2}$ | $V_{OC}$ = 0.16 V, $J_{SC}$ =1.04 nA·cm$^{-2}$ | Poor resistance response to pure CO$_2$ gas |
| B. Integrated light filter and solar cell | MIM resonator with PV cell [82] | Visible light, 235 W·m$^{-2}$ | / | $I$ from ~ 280 μA@5%RH, to ~220 μA@85%RH |
| | Pd nano-transducer with solar cell [83] | Visible light | $I_a$ = 300 μA | Response from ~1%@0.5%H$_2$, to ~3%@2% H$_2$, $t_{res}/t_{rec}$ ~ 100 s |
| C. Gas-sensitive heterojunction photovoltaics | Au@rGO/GaN(NRs)/Si [95] | 382 nm, 1.71 mW·cm$^{-2}$ | $I_a$ = 411.57 μA | Response ~38%@20 ppm CO |
| | Carbon/FMCPIB/TiO$_2$/FTO [96] | Fluorescent lamp | $V_{OC}$ = 0.63 V, $I_{SC}$ = 18.6 nA (≈35 μW·cm$^{-2}$) | Response ~8%@8 ppm NO$_2$, (1.3 μW·cm$^{-2}$) |
| D. Gas-sensitive lateral photovoltaics | S-hyperdoped silicon [99] | White light, 0.92 mW·cm$^{-2}$ | $V_{OC}$ = 0.6 mV, $I_{SC}$ ≈ 0.14 μA | Response 77.1%@50 ppm NH$_3$, $t_{res}/t_{rec}$ = 16/107 s |
| | N-hyperdoped silicon [101] | White light | $I_a$ ≈ 0.2 μA | Response 1654%@500ppm NO$_x$, $t_{res}/t_{rec}$ = 5/11 s |

[a] $V_{OC}$ = open-circuit voltage, $I_{SC}$($J_{SC}$)= short-circuit current (density), $I_a$ = current output in air.

Approach A relies on micro-processing technology in order to integrate a miniaturized gas sensor and a solar cell. Both components could perform different functions independently, but when they are combined, the solar cell provided the energy need for gas sensing. As to the device structure, gas sensor/insulator/solar cell tandem structures seem to be the best solution so far. Herein, the insulator ensures both components are connected in series, and the thin film gas sensor ensures that incident light could transmit through the upper layers. The choice of each component is relatively independent, because solar cells with high efficiency and gas sensors with high sensitivity are preferred for better device performance. Therefore, the challenges here are to develop solar cells with higher efficiency and gas sensors with better performance.

Approach B integrates a device (solar cell) and a function material (light filter). The light filter alone does not act as an independent device, because it has no transducing function, which is a major different from approach A with the gas sensor. The light filter senses the gas types and concentrations, changes its transmittance, and affects the efficiency of the solar cell beneath it. In order to get higher sensor sensitivity and selective response, the challenge here is to strictly match among the wavelength of the incident light, the change of transmittance spectrum of the light filter, and the absorbance spectrum of the solar cell.

Approach C uses rectifying junctions to realize PV effects, and the absorbent gas could alter the barrier heights of the junction between dissimilar materials to change the output signals. At least one material should be both semiconducting and gas sensitive, the latter meaning the electronic structures of the material can be modulated by a certain gas. By purposely choosing the other material, the baseline signal can be changed more flexibly. Since the change in electronic structures induced by gas adsorption occurs only in near surface areas, materials with small sizes and low dimensions are favorable, because the gas sensitivity is significantly affected by size effects. As mentioned before, further quantitative analysis of the sensing dynamics is highly anticipated and challenging.

Approach D uses lateral structure with uneven distribution of lights to realize PV effects, and the change of electronic structures of the materials directly reflects on the change in signals. The merit of this approach is that both functions can be realized based on a single material, which is beneficial for simplifying the structure and reducing the size of the device. Nevertheless, the lateral structure may require more length in one of the dimensions, so it is challenging to further reduce the size of the whole device.

### B. Exploitation of Applicable Materials

Materials are the very foundation for fabricating the sensing devices. Nowadays, new materials with novel properties are constantly being discovered. Many nanoscale and low-dimensional semiconductors have been applied to the concept of PV self-powered gas sensing in this review. Developing novel materials with more excellent optical/optoelectronic properties or gas sensing properties is a mainstream direction. Yet, the use of new materials does not always guarantee applicability. Instead, there are plenty of well-characterized photovoltaic and gas sensing materials as mentioned earlier. These have been proven efficient or applicable even long before the concept emerged. Therefore, the choice of applicable materials is vital to begin with.

However, studying the single function of sensing performance, optical characteristics and optoelectronic properties of a material is important for the use in gas sensor, light filter and solar cells, respectively. On the other hand, multi-functional materials or structures are more attractive, but they are more complex to study or design.

Since the materials aspect is focused for now, it should be noted that most semiconductors are inherently sensitive to certain types of gases or certain wavelengths or light at the same time, which is prerequisite for the use as multifunctional materials. The most typical multifunctional semiconductors in this review are silicon and perovskites. Nevertheless, gas adsorption induced photoresponse change was observed in SnO$_2$ [120], SnSe [121], SnS [122], PW$_{12}$-TiO$_2$ [123], MoS$_2$ [124], WS$_2$ [125], GaS [126], W$_{18}$O$_{49}$ [127], LaAlO$_3$/SrTiO$_3$ [128], ZnO [129]–[132], and so on, which were fabricated into field-effect transistors and metal-semiconductor-metal device structures. The literatures with different focuses commonly discuss how the absorbent gases affect the generation of a photocurrent. Therefore, after slight modifications these could be potential multifunctional materials used for the PV self-powered gas sensing.

### C. Integration of Multiple Functions

Rigorously, relying on artificial light sources like LEDs or fluorescent lamps does not truly mean zero power consumption in practical use. However, a steady light source is still required at this stage for stable output of sensing parameters. On the bright side, there are studies on the PV self-powered gas sensors which require only microwatts of light power or exhibit better performance under lower illumination level [96], [99]. Thus, integration of the gas sensor with the light source that possesses the most luminous efficiency at the least cost is one way to furthest reduce the power consumption and realize miniaturization at the same time. For example, Cho *et al.* [74] introduced an integration of gas-sensitive semiconductor metal oxide nanowires on micro light-emitting diodes ($\mu$LEDs) which showed excellent NO$_2$ sensitivity at the optimal operating power of $\sim$184 $\mu$W. Furthermore, Markiewicz *et al.* [133] reported a miniaturized sensing device integrating a photoactive material with a highly efficient LED light source, in which the total power consumption was as low as 30 $\mu$W. In fact, both sensors were passive types which did not take the power needed to measure the resistance change of the device into account. Despite that, the works provide a possible route towards an energy-efficient self-powered gas sensor integrated with the light source.

Regarding the means of energy utilization, multiple functions could be combined into a single chip for high-level integration. As introduced earlier, both gas sensing and photodetection functions were realized on a multifunctional hybrid device by integrating multiple units [78], [81]. Nevertheless, Reddeppa *et al.* [95] fabricated a solution-processed Au@rGO/GaN nanorods (NRs) hybrid structure with multi functions of self-powered photodetector and CO gas sensing (Fig. 8a). The optical property of the structure was enhanced by the surface plasmon resonance of Au nanoparticles (Fig. 8b). Working as a photodetector, the sensor was demonstrated to be responsive to the light illumination with wavelengths of 382 and 516 nm (Fig. 8c). Working as a self-powered gas sensor, it showed reversible response to CO gas from 5 to 50 ppm (Fig. 8d). These methods towards multifunctional devices will further reduce the required space for integration.

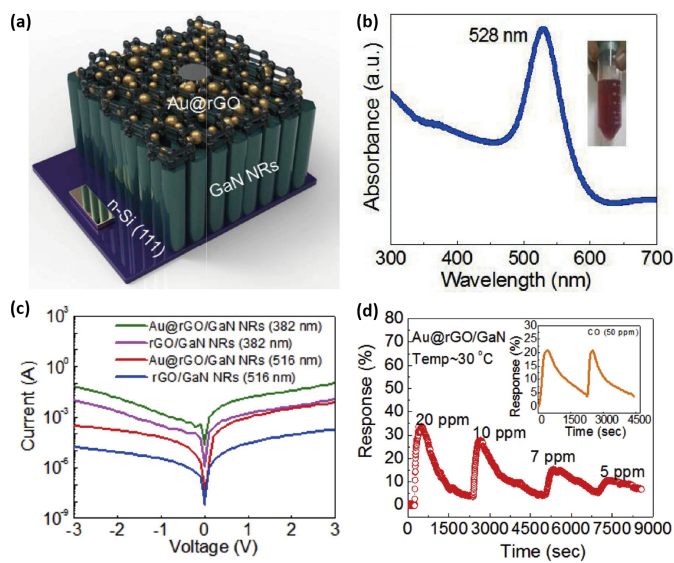

Fig. 8. Multifunctional Au@rGO/GaN NRs hybrid structure [95]. (a) Schematic representation of the structure. (b) UV–vis absorption spectrum of Au NPs in solution. The inset depicts the synthesized Au NPs in solution. (c) Photo illuminated currents of Au@rGO/GaN NRs and rGO/GaN NRs under 382 nm and 516 nm. (d) CO gas sensing properties of Au@rGO/GaN NRs at different gas concentrations at 30 °C (20 ppm, 10 ppm, 7 ppm, and 5 ppm). The inset depicts the CO gas sensing properties of rGO/GaN NRs at 50 ppm. *Reproduced with permission: 2019, Elsevier.*

Furthermore, there are many other ways to harvest ambient energy based on different mechanisms. Replacing solar cells with hybrid cells for simultaneously harvesting multi-type energies can be a solution to a more stable and sustainable energy source [134], [135]. Yun *et al.* [136] designed a ferroelectric BTO based multifunctional nanogenerator coupling the pyroelectric, photovoltaic, piezoelectric and triboelectric effects for simultaneously harnessing thermal, solar and mechanical energies in one structure. This work is prospective for maximizing energy conversion efficiency from ambient environment.

### D. Improvement of Sensing Performance

Bochenkov and Sergeev [35] proposed that an ideal chemical sensor would possess high sensitivity, dynamic range, selectivity and stability; low detection limit; good linearity; small hysteresis and response time; and long life cycle (although it is difficult to create an ideal sensor and only some of the parameters are important for specific application). It can be inferred from Table I that the current PV self-powered gas sensors are far from ideal. Measures have been taken to improve the gas sensing characteristics, such as surface modification for improved selectivity [86], [87], surface additives [93], and changing operating conditions like light intensity [91], [93], [99] or wavelength [96]. Readers may refer to the reviews on gas sensors for better guidance [34]–[48].

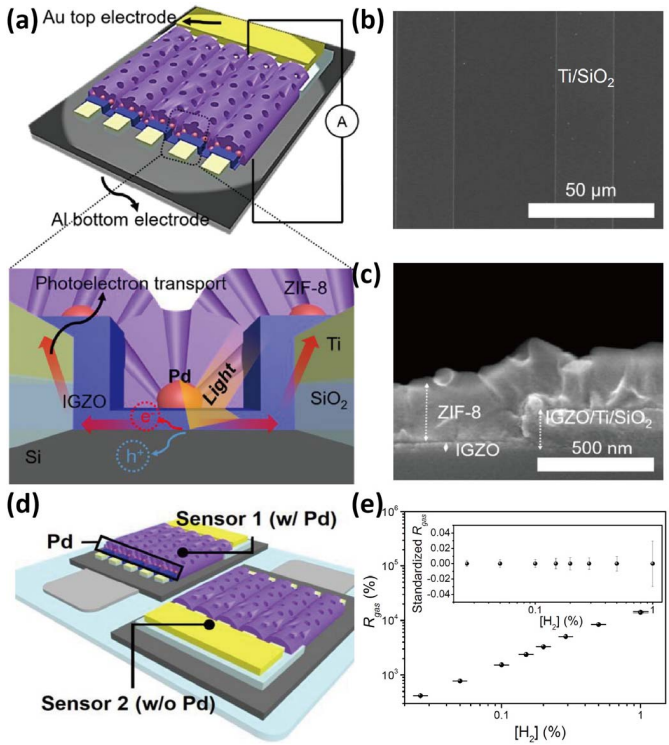

Fig. 9. Self-powered gas sensor based on ZIF-8/Pd-IGZO heterojunctions [137]. (a) Schematic of the ZIF-8/Pd-IGZO sensor fabricated on a p-type Si substrate with finger-type Ti/SiO2 electrodes. (b) Top-view and (c) cross-sectional SEM images of the sensor. (d) Sensing platform consisting of both ZIF-8/IGZO sensors with (sensor 1) and without (sensor 2) Pd catalysts. (e) Averaged response values under 15 different irradiances at each $H_2$ concentration. The inset shows the standardized result for each point. *Reproduced with permission: 2020, American Chemical Society.*

For a reliable PV self-powered gas sensor, it is important to reduce the dependence of sensing performance on light source. If possible, the sensing parameters should be insensitive to the change of the ambient light for practical purpose. D. Lee *et al.* [93] found that the relative change in photocurrent ($\Delta I_{sc}/I_{sc}$) of the Gr/Si/metal heterojunction was independent of the light intensity ($\sim 25–35\%@0–100$ mW·cm$^{-2}$) but sensitive to $H_2$ concentration ($0–\sim 60\%@0–1000$ ppm), which could be used for relatively reliable gas sensing. Besides exploring different material characteristics, turning to artificial light sources can guarantee a stable light input. However, using solar light or ambient light with arbitrary intensity is ideal to achieve zero power consumption. Kim *et al.* [137] proposed a PV self-powered hydrogen sensing platform having constant sensing responses regardless of light condition (Fig. 9a–c). The platform consisted of two photovoltaic units (Fig. 9d): (1) a Pd decorated n-IGZO/p-Si photodiode covered by a microporous zeolitic imidazolate framework-8 (ZIF-8) film (denoted sensor 1) and the same device configuration without the Pd catalyst as light intensity calibration module (denoted sensor 2). Thereinto, sensor 1 was sensitive to $H_2$ flux, whereas sensor 2 was not. The response values ($R_{gas}$) were calculated by $R_{gas}(\%) = 100 \cdot (I_1 - CI_2^\alpha)/CI_2\alpha$, where $I_1$ and $I_2$ were the photocurrents of sensor 1 and 2, respectively, and C and $\alpha$ were empirical constants. The sensor exhibited a uniform response with a standard deviation of 0.03 at 1% $H_2$ at most under 15 different irradiances (Fig. 9e). This indicates truly reliable and accurate gas sensing under unknown light intensity.

### E. Potential Applications

Different application needs are the driving force for the development of the concept. From a macro viewpoint, the development of wireless sensor networks (WSNs) leads to the omnipresent internet of things (IoTs) that will tremendously change our way of life. A WSN is a network of nodes that work in a cooperative way to sense and control the environment surrounding them [138]. Thereinto, the sensor nodes are essential for real-time monitoring and autonomous response. However, the major drawback of the WSNs is that they are battery dependent, so that they are designed to consume less operating energy which may cause security issues due to energy saving policies [138], [139]. Nevertheless, using the self-powered energy harvesters could greatly reduce the dependence on battery power [140], [141]. In scenarios like unattended air monitoring, some forms of energy sources like the mechanoelectrical energy may not as available as the photovoltaic energy, but low illumination indoor conditions could be challenging for powering the sensor nodes [139]. Therefore, developing PV self-powered gas sensors working under lower illumination conditions with high efficiencies can be a solution to these issues. Besides, solar powered unmanned aerial vehicles (UAV) equipped with sensor nodes can be a powerful tool to sense environmental gases in remote areas and survey large regions [142].

In fact, for other self-powered gas sensing systems used for different applications each energy source may not always available, which may cause serious reliable issues. Using solar or light energy by hybrid cells can be an important supplement for the self-powered energy source, although it can be varied by time, weather and so on. Anyway, the PV self-powered gas sensor will experience a long journey from concepts to real applications. Building a complete sensing system is inevitable for the practical use, which will always require the collaborative study in multidisciplinary fields.

### IV. CONCLUSION

In this review, recent advances towards the concept of the photovoltaic self-powered gas sensing are summarized and categorized as integrated gas sensor and solar cell, integrated light filter and solar cell, gas-sensitive heterojunction photovoltaics, and gas-sensitive lateral photovoltaics, respectively. These approaches are based on different materials, structures and mechanisms, and have their own characteristics. The concept has aroused wider attentions in the last decade and will certainly experience further development. Therefore, we put forward some prospects on future directions from materials to applications. Despite some progress in certain areas, it is still a long way from being practical. In our opinion, the PV self-powered gas sensor should be diverse in type, compact in size, high-efficient in energy harvesting and utilization, better performing and reliable in use. We hope practitioners of interests may be inspired after reading and jointly push the concept forwards.

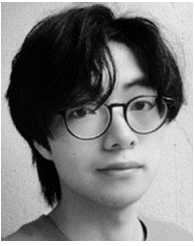

**Xiao-Long Liu** received the B.S. and B.A. degrees from the China University of Petroleum, Qingdao, in 2015, and the Ph.D. degree in optics from Fudan University, Shanghai, in 2020. As a Ph.D. student, his research was focused on fs-laser fabricated black silicon for gas sensors and photodetectors under the guidance of Prof. Jun Zhuang. He is a Postdoctoral Researcher with the FemtoBlack Project in Aalto University, Finland.

**Yang Zhao**, photograph and biography not available at the time of publication.

**Wen-Jing Wang**, photograph and biography not available at the time of publication.

**Sheng-Xiang Ma**, photograph and biography not available at the time of publication.

**Xi-Jing Ning**, photograph and biography not available at the time of publication.

**Li Zhao**, photograph and biography not available at the time of publication.

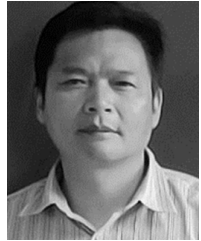

**Jun Zhuang** received the Ph.D. degree from the Shanghai Institute of Optics and Fine Mechanics in 1996. He is a Professor with the Department of Optical Science and Engineering, Fudan University, Shanghai. His research interests include the properties of functional materials, and their applications in optoelectronics, photovoltaics, and gas sensing.